\documentclass[fleqn, usenatbib]{mnras}
\usepackage{newtxtext,newtxmath}
\usepackage[T1]{fontenc}
\usepackage{ae,aecompl}

\usepackage{graphicx}	
\usepackage{amsmath}	
\usepackage{amssymb}	
\usepackage{fixltx2e} 

\newcommand\altaffilmark[1]{$^{#1}$}
\newcommand\altaffiltext[1]{$^{#1}$}
\newcommand{\angstrom}{\mbox{\normalfont\AA}}

\title[Binarity-Driven Bias]{Signatures of unresolved binaries in stellar spectra: implications for spectral fitting}

\author[El-Badry et al.]{
\parbox[t]{\textwidth}{ 
Kareem El-Badry\thanks{E-mail: kelbadry@berkeley.edu}\altaffilmark{1,2},
Hans-Walter Rix\altaffilmark{2},
Yuan-Sen Ting\altaffilmark{3}, 
Daniel R. Weisz\altaffilmark{1}, 
Maria Bergemann\altaffilmark{2},
Phillip Cargile\altaffilmark{5},
Charlie Conroy\altaffilmark{5},
and Anna-Christina Eilers\altaffilmark{2,6}
} 
\vspace*{10pt} \\
\altaffiltext{1}{Department of Astronomy and Theoretical Astrophysics, University of California Berkeley, Berkeley, CA 94720, USA} \\
\altaffiltext{2}{Max Planck Institute for Astronomy, D-69117 Heidelberg, Germany} \\
\altaffiltext{3}{Research School of Astronomy \& Astrophysics, The Australian National University, Canberra ACT 0200, Australia} \\
\altaffiltext{5}{Department of Astronomy, Harvard University, Cambridge, MA, 02138, USA} \\
\altaffiltext{6}{International Max Planck Research School for Astronomy \& Cosmic Physics at the University of Heidelberg, Germany}
}

\date{Accepted 2017 October 18. Received 2017 October 17; in original form 2017 September 11}

\pubyear{2017}

\begin{document}
\label{firstpage}
\pagerange{\pageref{firstpage}--\pageref{lastpage}}
\maketitle

\begin{abstract}
The observable spectrum of an unresolved binary star system is a superposition of two single-star spectra. Even without a detectable velocity offset between the two stellar components, the combined spectrum of a binary system is in general different from that of either component, and fitting it with single-star models may yield inaccurate stellar parameters and abundances. We perform simple experiments with synthetic spectra to investigate the effect of unresolved main-sequence binaries on spectral fitting, modeling spectra similar to those collected by the APOGEE, GALAH, and LAMOST surveys. We find that fitting unresolved binaries with single-star models introduces systematic biases in the derived stellar parameters and abundances that are modest but certainly not negligible, with typical systematic errors of $300\,\rm K$ in $T_{\rm eff}$, 0.1 dex in $\log g$, and 0.1 dex in $[\rm Fe/H]$ for APOGEE-like spectra of solar-type stars. These biases are smaller for spectra at optical wavelengths than in the near-infrared. We show that biases can be corrected by fitting spectra with a binary model, which adds only two labels to the fit and includes single-star models as a special case. Our model provides a promising new method to constrain the Galactic binary population, including systems with single-epoch spectra and no detectable velocity offset between the two stars.
\end{abstract}

\begin{keywords}
binaries: spectroscopic --- stars: abundances --- Galaxy: stellar content --- methods: data analysis
\end{keywords}

\section{Introduction}
Roughly half of all solar-type stars reside in binary systems \citep{Raghavan_2010, Moe_2017}. Outside the solar neighborhood ($d\gtrsim100\,{\rm pc}$), most binaries are spatially unresolved and will fall within the same fiber in spectroscopic surveys. Short-period binaries are easily identified spectroscopically by a line-of-sight velocity offset between the two components \citep{Pourbaix_2004, Matijevic_2010, Fernandez_2017, Merle_2017} or time-variable radial velocity measurements \citep{Troup_2016, Gao_2017}. However, most solar-type binary systems have orbital periods of hundreds to thousands of years (see \citealt{Duchene_2013}, for a review) and are missed by binarity detection methods based on the Doppler shift.

Absent a detectable velocity offset, the spectrum of an unresolved binary still contains signatures of the unseen companion. These are strongest when the two components have very different spectral features, as is the case with white dwarf -- main sequence binaries \citep{RebassaMansergas_2007, Ren_2013}. However, the presence of an unresolved companion changes the observable spectrum in \textit{all} cases where the two stars are not identical.

This is on one hand useful for the study of binary systems, as it implies that many binaries should be detectable from detailed modeling of their spectra, even without a detectable velocity offset. However, it also presents a challenge for Galactic surveys aiming to measure the properties of stars by fitting their spectra. When the spectrum of an unresolved binary is fit with single-star models, contamination from the unseen companion may introduce biases in the inferred atmospheric parameters and abundances. Quantifying and remedying these biases is the subject of this paper. 

\section{Methods}
\label{sec:methods}

In order to disentangle the effects of unresolved binaries from systematic errors in stellar models, we conduct our analysis with synthetic spectra. 

We generate spectra for single stars using 1D-LTE atmosphere models computed by the \texttt{ATLAS12} code \citep{Kurucz_1970, Kurucz_1979, Kurucz_1992}. Radiative transfer calculations are carried out with \texttt{SYNTHE} \citep{Kurucz_1993}. We vary (and later fit for) four stellar labels: $T_{\rm eff},\,\log g,\,[\rm Fe/H],$ and $[\rm \alpha/Fe]$, where Mg, Si, S, Ar, Ca, and Ti are included in $\alpha$ and are varied in lockstep. 

To efficiently calculate the synthetic spectrum at an arbitrary point in label space (as is required during fitting) we interpolate between nearby models at each wavelength pixel using a neural network. This approach, which makes it possible to self-consistently fit for many stellar labels simultaneously, is similar to that developed in \citet{Ting_2016} and \citet{Rix_2016}. It was introduced in \citet{Ting_2017b} and will be fully explained in Ting et al. (2017c, in prep). In brief, we first generate an irregular grid of ab-initio Kurucz model spectra throughout the region of label space within which a spectral model is desired. These spectra are then used as a training set to train an individual neural network to predict the flux at each wavelength pixel as a function of labels. The full spectral model then consists of all the individual neural networks for all wavelength pixels stitched together. We have verified through cross-validation that typical interpolation errors in model fluxes are small ($\ll $ 1 percent). We produce synthetic spectra with the wavelength coverage and typical S/N per pixel of three surveys: 
\begin{itemize}
\item APOGEE, with $R=22,500$, $S/N=100$, and coverage across $1.51-1.70 \rm\,\mu m$, with some gaps \citep{Holtzman_2015}.
\item GALAH, with $R=28,000$, $S/N=100$, and coverage over four $\sim 250\,\angstrom$ channels between $0.47\,\mu{\rm m}$ and $0.79\,\mu{\rm m}$ \citep{DeSilva_2015}.
\item LAMOST, with $R=1,800$, $S/N=30$, and coverage across $0.37-0.90\,\mu {\rm m}$ \citep{Cui_2012}.
\end{itemize}
These surveys were chosen to span a range of resolution, $S/N$, and wavelength coverage representative of modern spectroscopic surveys. A primary goal of all three surveys is to derive precise stellar parameters and abundances for large samples of stars in order to study the formation and enrichment history of the Galaxy.

Synthetic spectra are calculated at $R=300,000$, degraded to lower resolutions assuming a Gaussian line spread function (LSF) with ${\rm FWHM} = \lambda/R$, and resampled on the wavelength grid of each survey. Best-fit labels are determined through full-spectrum fitting with $\chi^2$ minimization on normalized spectra. We normalize APOGEE-like spectra using the \textit{Cannon}-type continuum normalization routine from the \texttt{APOGEE} package \citep{Bovy_2016}. For LAMOST-and GALAH-like spectra, we normalize by dividing the spectrum by a smoothed version of itself, as in \citet{Ho_2017} and \citet{Ting_2017b}. For our idealized experiments, the choice of normalization method has a negligible effect on the best-fit stellar parameters. We emphasize that we make no attempt to closely reproduce the LSF, continuum normalization, or fitting procedure of the surveys we emulate. Rather, we seek to qualitatively measure how unresolved binarity changes stellar spectra at different resolutions and wavelengths, without making strict assumptions about implementation. 

We have verified that our approach can recover the labels of single-star spectra with added noise with high fidelity. We also experimented with generating and fitting spectra using NLTE spectra computed using the 1D hydrostatic MAFAGS-OS model atmospheres \citep{Grupp_2004, Bergemann_2012} instead of \texttt{ATLAS12} and \texttt{SYNTHE}, obtaining very similar results for the effects of unresolved binarity. 

We emphasize that the choice of spectral model and normalization method have negligible effects on the best-fit stellar parameters for each survey only because the spectral model and the spectra to be fit are always generated self-consistently. For example, rather than predicting spectra normalized with the ``true'' continuum, the model for normalized spectra predicts spectra normalized in the same way as the spectra to be fit (for each mock-survey). In cases where there is a mismatch between the model and the spectra being fit, changes to the normalization procedure and other details of the spectral model will have more significant effects.

\section{Fitting binary spectra with single-star models}
\label{sec:fitting_with_single}

\begin{figure*}
	\includegraphics[width=\textwidth]{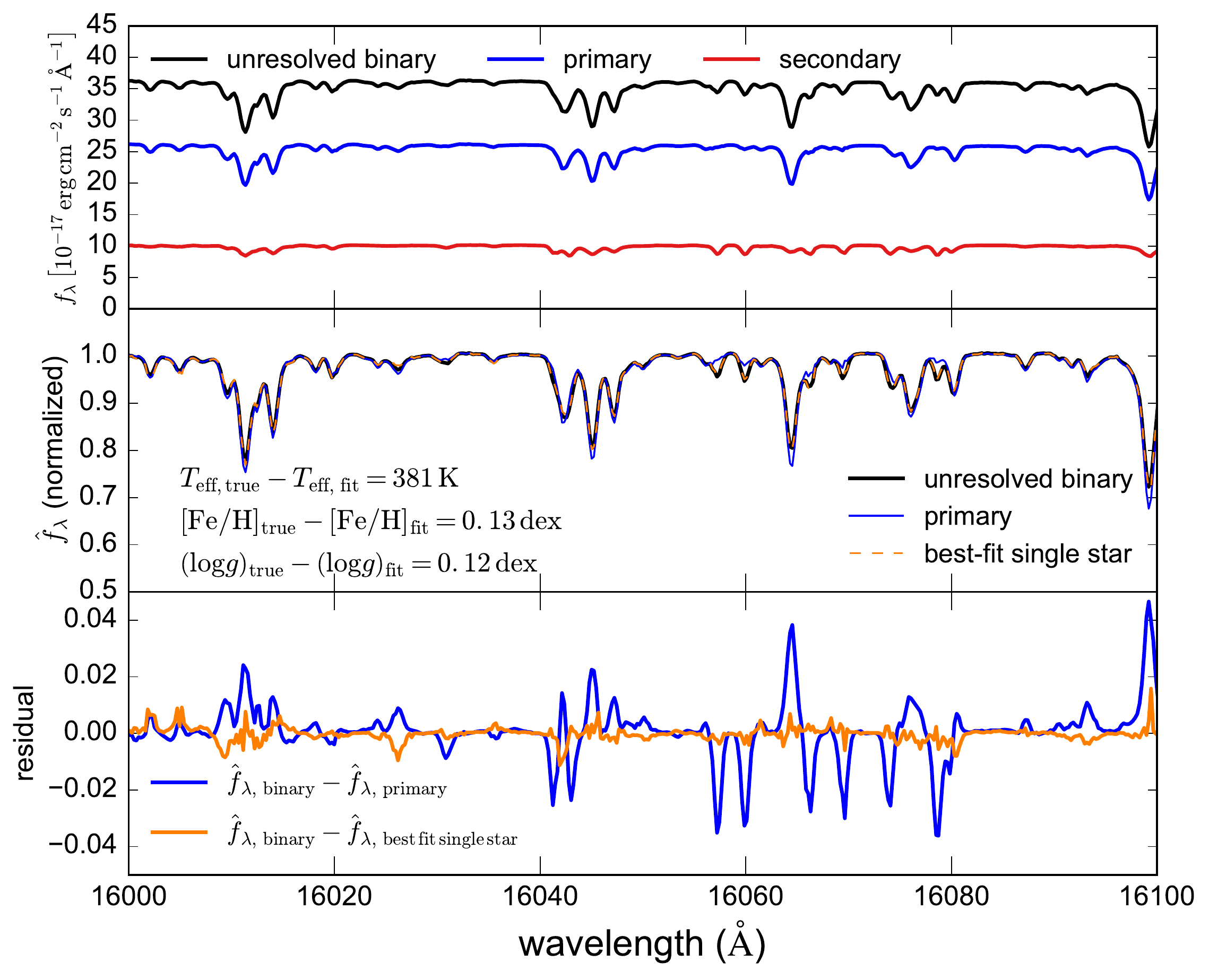}
    \caption{Effect of an unresolved companion on the observable APOGEE-like spectrum of a K-dwarf star. \textbf{Top}: synthetic spectrum of the primary, secondary (mass ratio $q=m_2/m_1 = 0.75$), and their sum, with no velocity offset. \textbf{Middle:} normalized spectrum of the primary star, unresolved binary, and the best-fit single-star model. \textbf{Bottom}: residuals of the spectra of the true primary star and the best-fit single star model with respect to the unresolved binary spectrum. The unresolved binary spectrum deviates subtly but nontrivially from the that of the primary (blue). It can be well-fit by a single-star model (orange), but the inferred stellar labels are biased toward lower temperature and metallicity.} 
    \label{fig:schematic_figure}
\end{figure*}

Figure~\ref{fig:schematic_figure} illustrates the effects of unresolved binarity on a small wavelength range of an APOGEE-like spectrum for a typical K-dwarf star ($T_{\rm eff} = 5000\,{\rm K},\,\log g = 4.59,\,[\rm Fe/H] = -0.2$). The top panel shows the spectrum of the primary (blue), secondary (red), and their sum (black). We always assume that the secondary has the same age and abundances as the primary (neglecting the subtle effects of atomic diffusion; see \citealt{Dotter_2017}), so that its spectrum is completely determined by the mass ratio $q=m_2/m_1$. Atmospheric parameters of the secondary ($T_{{\rm eff}}=4100\,{\rm K},\,\log g=4.72$ for the star in Figure~\ref{fig:schematic_figure}) are calculated for a given mass ratio using \texttt{MIST} isochrones \citep{Choi_2016}. 

In the middle panel, we show the continuum-normalized spectrum of the binary and primary star, as well as the result of fitting the binary spectrum with a single-star model. The corresponding residuals are shown in the bottom panel. Because the primary and secondary stars have different temperatures, they have noticeably different spectral features. For example, the absorption line at $16064\,\angstrom$ is weaker in the secondary, while the lines at $16057\,\angstrom$ and $16060\,\angstrom$ are much weaker in the primary. As a result, the normalized binary spectrum deviates from that of the primary in several lines. 

Fitting the unresolved binary spectrum with a single-star model produces a good fit, with significantly smaller residuals compared to the primary star spectrum. However, the labels of the best-fit single-star model differ significantly from the true labels of the primary. This case study shows that fitting a binary spectrum with a single-star model can introduce biases in the derived stellar parameters and abundances. We now investigate these systematically.

\begin{figure*}
\includegraphics[width=\textwidth]{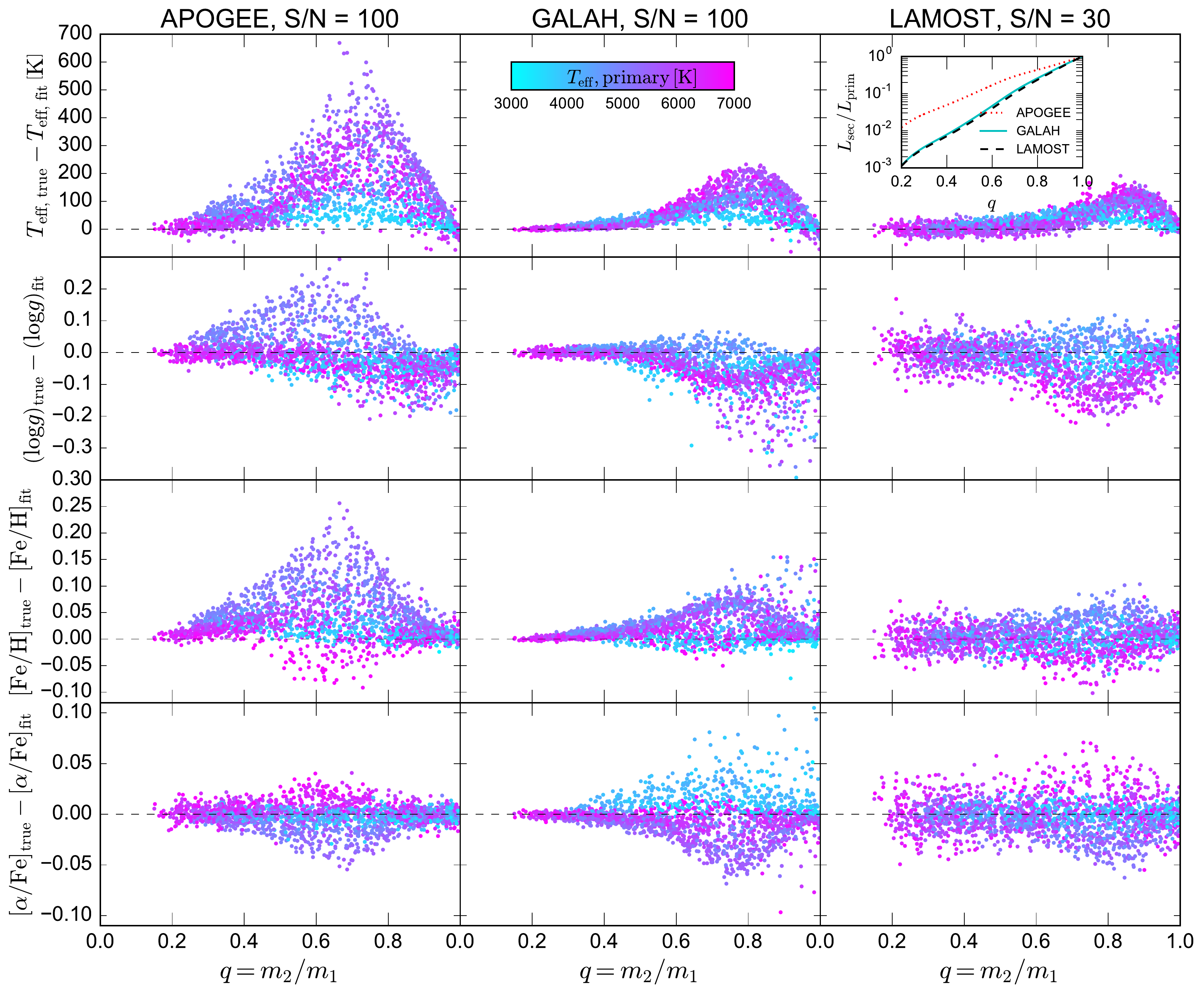}
    \caption{Effects of unresolved binarity on the accuracy of stellar labels recovered from spectral fitting. We model 2000 synthetic unresolved main-sequence binaries with a realistic distribution of orbits, mass ratios, and stellar parameters. We produce synthetic spectra with wavelength coverage, resolution, and signal-to-noise per pixel similar to three spectroscopic surveys (Section~\ref{sec:methods}) and fit them with single-star models. Unresolved binarity introduces systematic biases in the inferred stellar parameters which are largest at $q\approx 0.7$. Inset shows the luminosity ratio at each survey's wavelength coverage for a Sun-like primary. Binarity leads to larger systematics for APOGEE-like spectra than for optical spectra because lower-mass companions contribute a larger fraction of the total light in the near-infrared.} 
    \label{fig:binarity_bias}
\end{figure*}

\begin{figure*}
\includegraphics[width=\textwidth]{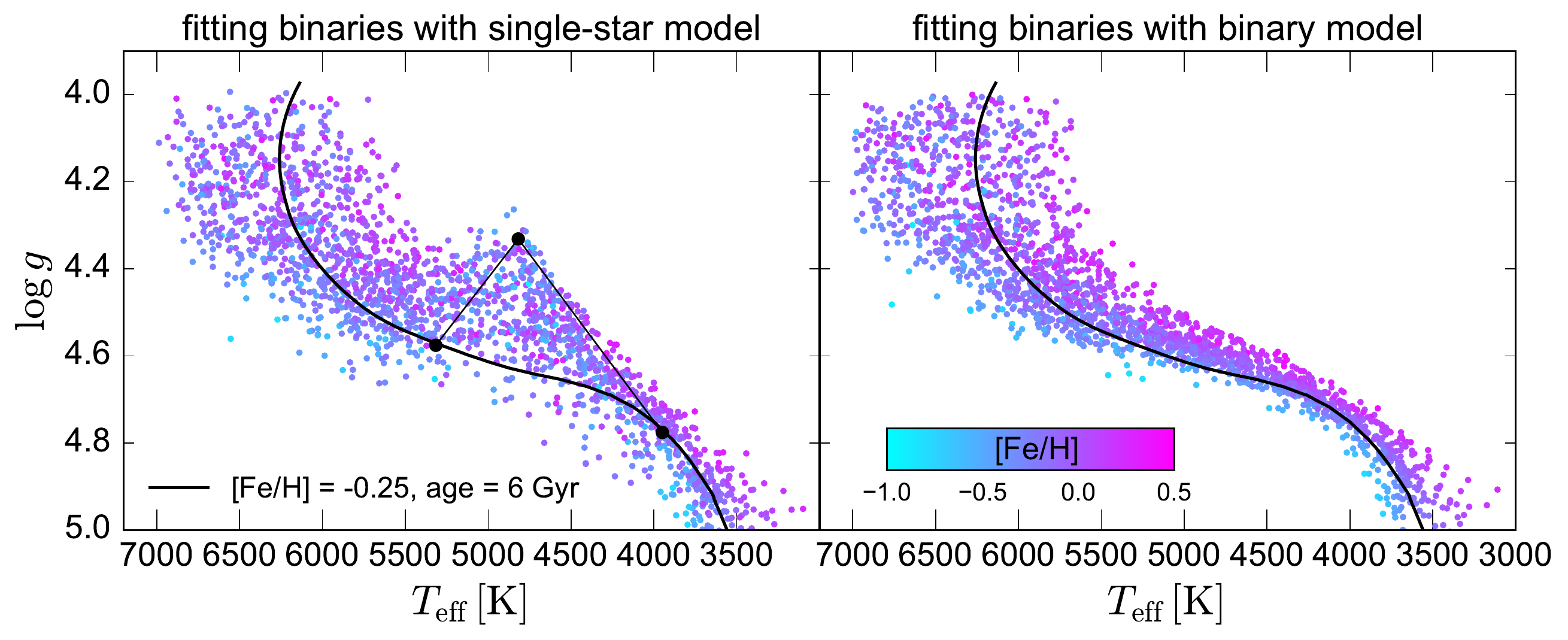}
\caption{Effect of unrecognized binarity on the HR diagram for APOGEE-like spectra. \textbf{Left}: We generate synthetic spectra for a population of 2000 unresolved main-sequence binaries and fit them with single-star models, plotting the best-fit parameters. A \texttt{MIST} isochrone is shown for comparison. For one system, black points show the \textit{true} $T_{\rm eff}$ and $\log g$ of the primary and secondary star, as well as the best-fit values obtained by fitting the  binary spectrum with a single-star model. Unrecognized binarity causes the inferred stellar parameters of many binaries to scatter above the main sequence and disrupts the monotonic decrease in $T_{\rm eff}$ with metallicity. \textbf{Right}: Best-fit parameters for the primary found by fitting spectra with a binary model (see Section~\ref{sec:bin_model}). Fitting a binary model recovers the true parameters without systematic biases.}
\label{fig:hrd}
\end{figure*}

We construct mock spectra for 2000 unresolved binaries by combining pairs of main-sequence single stars with masses spanning $(0.2-1.7)\,M_{\odot}$ and ages spanning $(1 - 12)\,\rm Gyr$, with ${3100\,\rm K} \leq T_{\rm eff} \leq {7000\,\rm K}$ and ${4.0\,\rm dex} \leq \log g \leq {5.05\,\rm dex}$. Masses for both the primary and secondary are drawn from a uniform distribution. Atmospheric parameters for both stars are calculated from \texttt{MIST} isochrones; cases where either star has left the main sequence are discarded. We assign metallicities and $\alpha-$abundances to each pair simultaneously by drawing randomly from labels of stars in APOGEE DR13 \citep{Holtzman_2015}. We draw orbital periods from the lognormal period distribution for solar-type stars found in \citet{Duchene_2013}; we assume random orbital orientations and phases to calculate line-of-sight velocity offsets for each system, assuming circular orbits. We exclude the $\sim 20\% $ of orbital configurations with velocity offsets $\left|\Delta v\right| > 10\,\rm km\,s^{-1}$, as these could potentially be identified as double-line binaries in high-resolution spectra and removed from the sample. We then sum the unnormalized single-star spectra, add Gaussian noise to bring the binary spectra to the fiducial S/N of each survey, and fit them with single-star models. The results of this experiment are shown in Figure~\ref{fig:binarity_bias}. For each mock-survey, we compare the true labels of the primary star to those found by fitting a single-star model to the binary spectrum. 

In the limit of low $q$, systematic errors in labels are small, because the secondary star contributes a negligible fraction of the total light. At $0.4\lesssim q \lesssim 0.9$, fitting a single-star model leads to a best-fit $T_{\rm eff}$ that is on average too low, because the primary spectrum is contaminated by a cooler companion. For all surveys, systematic errors in all labels except $\log g$ are largest at $0.6 \lesssim q \lesssim 0.9$. The mass ratio at which binarity has the largest effect is somewhat lower for APOGEE-like spectra than for LAMOST- and GALAH-like spectra, because cooler secondaries contribute more significantly to the total spectrum in the near-infrared than at optical wavelengths.
The biases caused by unresolved binarity are subtle but not negligible. For example, for APOGEE-like spectra of solar-type stars with $0.5<q<0.8$, the median errors in $T_{\rm eff}$ and $[\rm Fe/H]$ are 330 K and 0.1 dex, respectively. 
In the limit of $q\sim 1$, the two components have identical spectra. However, since we include velocity offsets of up to $10\,\rm km\,s^{-1}$ (with a median $\left|\Delta v\right|= 1.2\,\rm km\,s^{-1}$), some biases remain at $q\sim 1$. The largest errors in recovered parameters for $q\sim 1$ binaries are in $\log g$, which is most sensitive to changes in line shape. We have verified that biases at $q\sim 1$ disappear when $\Delta v=0$. 

The strength of the bias in best-fit single star labels varies with the temperature of the primary: systems with low $T_{\rm eff}$ have smaller systematic offsets on average. This is because $T_{\rm eff}$ is almost constant with mass along isochrones low on the main-sequence. In binaries where $T_{\rm eff,\,primary} \lesssim 4000\,\rm K$, the binary and secondary have similar temperatures, and thus, similar spectra. 

The typical systematic biases at intermediate $q$ are larger in APOGEE-like spectra than in GALAH- or LAMOST-like spectra due to APOGEE's redder wavelength coverage. Lower-mass stars are always redder at fixed metallicity, and they therefore contribute a larger fraction of the total light at longer wavelengths. This is illustrated in the inset in the upper-right panel of Figure~\ref{fig:binarity_bias}, which shows as a function of $q$ the luminosity ratio of the secondary star to the primary, integrated over the wavelength coverage of each survey. At $q=0.7$, the secondary contributes $\sim 30\%$ as much light as the primary at APOGEE wavelengths, compared to only $\sim 12 \%$ and $\sim 11\%$ at the wavelengths probed by GALAH and LAMOST, respectively. Likely for this reason, previous work has found the biases introduced by unresolved binaries to be relatively small at optical wavelengths \citep{Matijevic_2010}.

Here we only model main-sequence binaries, as binarity will likely have a much smaller effect on stellar parameter estimates for red giant stars. This is because a dwarf secondary will contribute a minuscule fraction of the total light in giant-dwarf binaries, while in giant-giant binaries, the two components will necessarily have very similar masses, and consequently, similar spectra. 

In Figure~\ref{fig:hrd}, we show how unresolved binarity changes the distribution of stars in the best-fit $T_{\rm eff} - \log g$ plane for APOGEE-like spectra. The left panel shows the results of fitting the spectra of the binary population from Figure~\ref{fig:binarity_bias} with single-star models. For most binaries, the best-fit single-star parameters scatter off the main sequence. As one might expect, the spectral contributions of a cooler companion bias the inferred $T_{\rm eff}$ to lower temperatures than the true value for the primary. The best-fit $\log g$ is often lower than the true value for either star. This occurs because $\log g$ is significantly covariant with both $T_{\rm eff}$ and $\rm [Fe/H]$ \citep[see e.g.][]{Ting_2017}, and the single-star model can best accommodate the change in the spectral line profiles due to the cooler star by decreasing both $T_{\rm eff}$ and $\log g$. As a result, single-star fits to binary spectra often lie above the main sequence. This suggests that unresolved binarity may be partially responsible for the presence of ``main sequence upturn'' stars found in many spectroscopic surveys \citep[e.g][]{Kordopatis_2013, Sharma_2017}, though limitations in atmosphere models likely also play a role. The right panel of Figure~\ref{fig:hrd} shows that fitting spectra with a binary model, as described in the next section, can correct these biases.

\section{Fitting a binary spectral model}
\label{sec:bin_model}

We now demonstrate how the biases discussed in the previous section can be corrected by fitting a binary model; i.e., fitting the spectrum with a sum of two single-star spectra. We construct a binary spectral model by adding two additional labels to the single-star model: the mass ratio $q$, which sets $T_{\rm eff}$ and $\log g$ of the secondary from isochrones, and the line-of-sight velocity offset $\Delta v$. We note that the assumption that the primary and secondary star reside on a theoretical isochrone may be problematic for cool stars. There is in principle no need to enforce this restriction; fitting all parameters of the primary and secondary simultaneously simply comes at the cost of greater model complexity.

\begin{figure}
\includegraphics[width=\columnwidth]{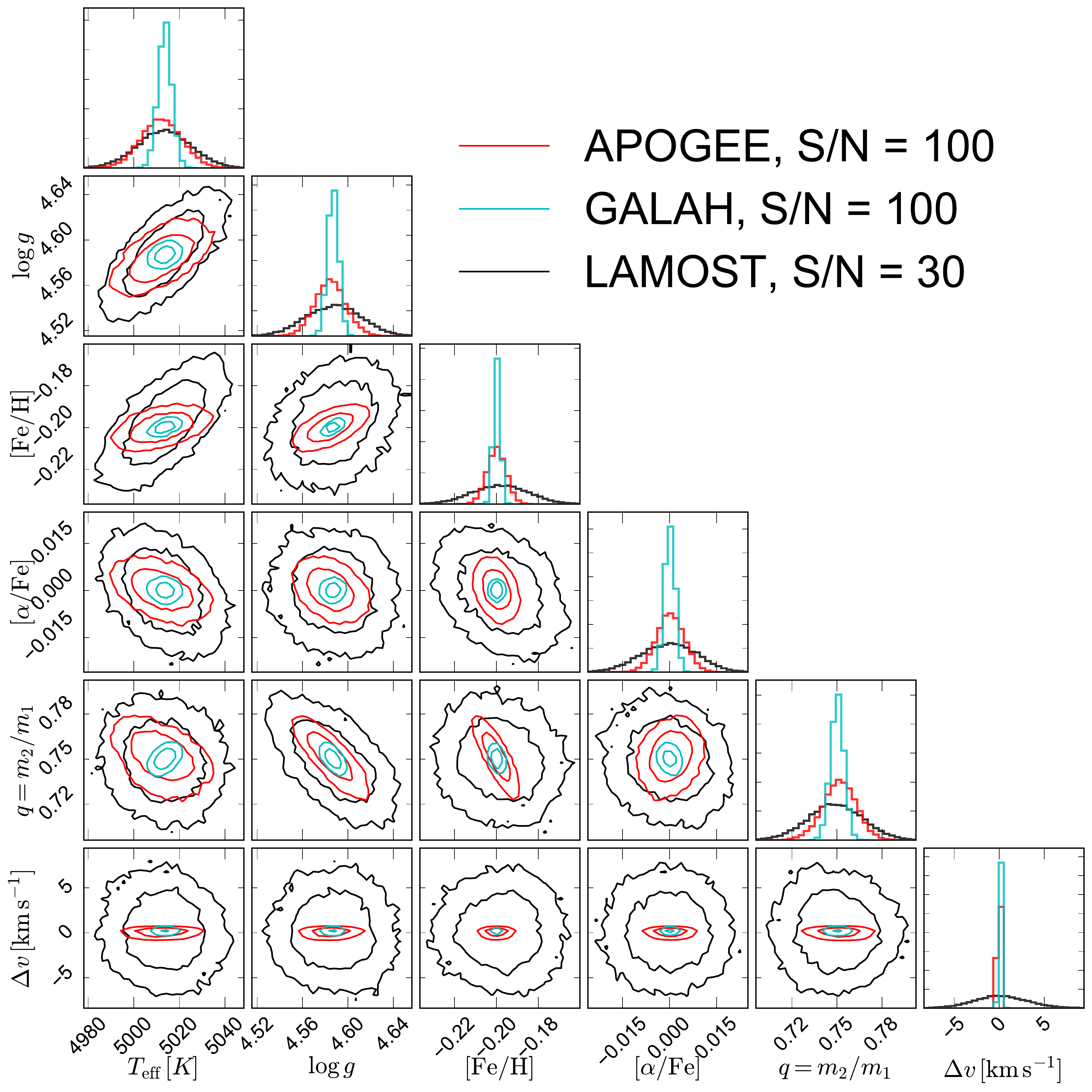}
\caption{68 and 95\% probability contours obtained by fitting a binary model to the synthetic spectrum of an unresolved binary as it would be observed by three spectroscopic surveys. Fitting a binary model corrects the bias that is present when the spectrum is fit by a single-star model and adds only two additional labels: the mass ratio, $q$, and the velocity offset, $\Delta v$. A binary model makes it possible to identify unresolved binaries even in systems where no velocity offset is measurable.}
\label{fig:mcmc_comparison}
\end{figure}

In Figure~\ref{fig:mcmc_comparison}, we show the results of fitting the synthetic spectrum of the binary shown in Figure~\ref{fig:schematic_figure}, as it would be observed by each survey, with a binary model. The posterior for each spectrum was sampled with \texttt{emcee} \citep{ForemanMackey_2013} with non-informative flat priors, and marginalized projections are generated with \texttt{corner} \citep{corner}; constraints on $\Delta v$ are marginalized over the uncertainties in the velocities of both stars, which are fit independently. Unsurprisingly (since the data are generated and fit with the same model), the binary model is able to recover the true input parameters without systematic biases. The constraining power of spectra from the three surveys is in large part a function of S/N and pixel count -- so LAMOST-like spectra provide weaker constraints for most labels than APOGEE- and GALAH-like spectra -- but it also depends on the sensitivity of different parts of the spectrum to different labels. Thus, the covariances between different labels in Figure~\ref{fig:mcmc_comparison} are somewhat different for different surveys. 

APOGEE- and GALAH like spectra have comparable resolution and adopted $S/N$, but constraints on most labels are stronger for GALAH-like spectra, in large part due to the higher spectral information content at short wavelengths, where more lines contribute in fixed-length wavelength interval. For the same reason, constraints on $T_{\rm eff}$ and $\log g$ for LAMOST-like spectra are nearly as strong as those from APOGEE-like spectra despite the significantly lower adopted $S/N$.
The 68 and 95\% probability contours show that, at the fiducial S/N assumed in our analysis, the surveys we consider are sufficiently informative to constrain the parameters of a binary model with high fidelity (at least at $q\sim 0.75$). The additional labels introduced by the binary model ($q$ and $\Delta v$) are not strongly convariant with other labels, though all three surveys show a negative covariance between $q$ and $\log g$; i.e., decreasing $q$ increases $\log g$ of the secondary, which has a similar effect on the total spectrum as increasing $\log g$ of the primary.

\begin{figure}
\includegraphics[width=\columnwidth]{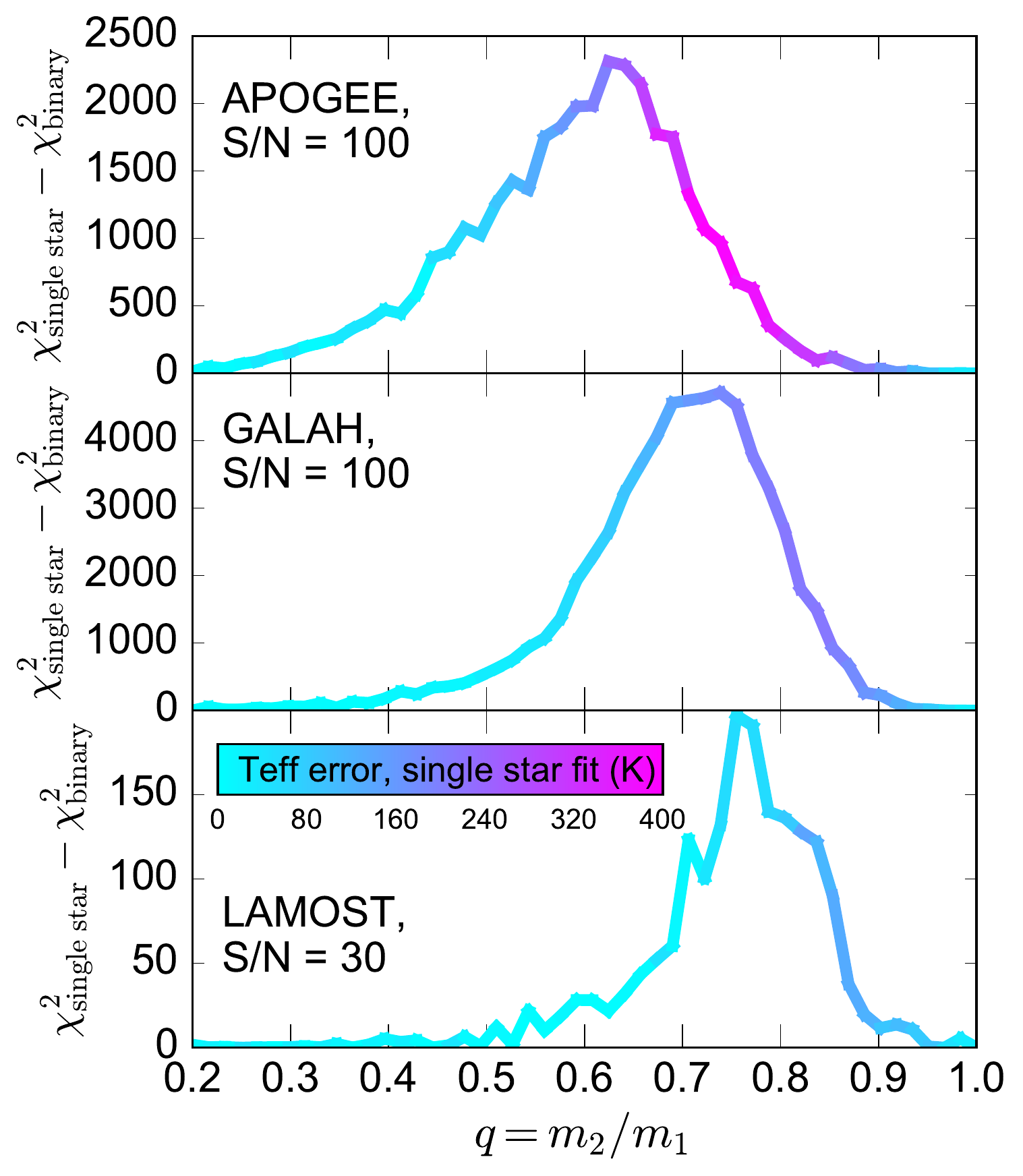}
\caption{Difference in total $\chi^2$ when fitting the synthetic spectrum of an unresolved binary with a Sun-like primary with a single-star model and a binary model (i.e., a sum of two single-star models). At intermediate mass ratios, unresolved binaries can be identified as systems which are significantly better fit by a binary model than a single-star model. } 
\label{fig:chi2_diff}
\end{figure}

This fitting approach can be used to constrain the Galactic binary population: probable binaries can in principle be identified as stars which are significantly better-fit in a $\chi^2$ sense by a binary model than by a single star. An important question is whether binary spectra can reliably be distinguished from single-star spectra with different parameters. We investigate this in Figure~\ref{fig:chi2_diff}. 

For a Sun-like primary star, we generate synthetic binary spectra with $q$ varying from 0.2 to 1. We set $\Delta v = 0$, since this is the limit in which detecting binarity is most challenging. We then fit each spectrum with both a single-star and a binary model. The difference in $\chi^2$ values between these fits quantifies ``how much better'' the spectrum is fit with a binary model. At intermediate mass ratios, the $\chi^2$ difference between the best-fit binary and single-star models is significant: $\Delta \chi^2 \gtrsim 100$ for LAMOST-like spectra, and $\Delta \chi^2 \gtrsim 1000$ for APOGEE- and GALAH-like spectra. The $q$ range where binaries are detectable is similar (though not identical) to the range in which fitting a single-star model introduces the largest systematics. 

We find that fitting a single-star spectrum with a binary model always converges to $q=1$ and $\Delta v = 0$, (which is completely degenerate with the single-star case) and recovers the correct labels for the primary star, with $\Delta \chi^2$ of order unity. Cases where the binary model provides a better fit but $\Delta \chi^2$ is small should be interpreted as consistent with being single stars. It it thus straightforward (in the limit of perfect models) to detect binaries with $0.6\lesssim q \lesssim 0.85$ in LAMOST- and GALAH-like spectra; in APOGEE-like spectra, binaries are in principle detectable at mass ratios as low as $q\sim 0.3$.

These experiments are of course idealized, as they assume no systematic shortcomings in the model spectra. Detecting unresolved binaries in real spectra will present challenges which do not arise with synthetic data. For example, systematic errors in stellar models may cause some single star spectra to be better-fit by a binary model. In this case, care must be taken in formulating a heuristic to determine what $\Delta \chi^2$ constitutes reliable evidence of binarity. A useful approach for fitting real spectra may be to use a data-driven model like \textit{The Cannon} \citep{Ness_2015} as opposed to synthetic spectra. Another complication not considered in this work is the effect of stellar rotation, which can broaden spectral lines in a manner superficially similar to a velocity offset in a binary system with $q\sim 1$. This effect is small for most cool stars but becomes important at $T_{\rm eff} \gtrsim 6500\,\rm K$ \citep{Glebocki_2000}. Because the effective kernel for rotational broadening changes spectral line profiles differently from a velocity offset in a binary system, a straightforward solution is to incorporate the projected rotation velocity as another label in the spectral model.

\section{Conclusions}
\label{sec:discussion}
We have shown that unrecognized binarity in main-sequence stars can introduce biases in stellar labels found by fitting single-star models. These can be significant compared to the target measurement precision of high-resolution spectroscopic surveys (which is $\sigma_{{\rm Teff}}\sim 50\,{\rm K,}\,\,\sigma_{{\rm [Fe/H]}}\sim0.05$ dex for APOGEE and GALAH) and present a potential challenge for works requiring precise measurements of stellar parameters and abundances of main-sequence stars, e.g., for chemical tagging \citep{Freeman_2002, Ting_2015, Hogg_2016, Ting_2016a}. These biases can be eliminated by fitting a binary model, with a modest increase in complexity. 

Fitting a binary model also makes it possible to spectroscopically identify main-sequence binary systems, even absent a velocity offset between the two stars: if the two components of a binary system have different temperatures, their combined spectrum will be better-fit by a binary model than a single-star model. The spectral signatures of binarity are stronger in the near-infrared than at optical wavelengths, because a cooler secondary contributes a larger fraction of the total light at longer wavelengths. For APOGEE-like spectra, the signatures of binarity are strongest for systems with $0.4\lesssim q \lesssim 0.8$. We will present results of fitting APOGEE spectra with a data-driven binary model in forthcoming work. 

An important limitation of fitting a binary model is that, without distance information, our method cannot detect binaries with mass ratios near $q=1$ and small velocity offsets: because the two components of a $q=1$ binary system have essentially identical spectra, their normalized combined spectrum will be indistinguishable from that of a single star. However, in this case there is also no bias in the inferred stellar parameters and abundances. The degeneracy between a single star and a $q=1$ binary can also be broken with moderate-accuracy distance information, since at fixed distance, a $q=1$ binary will be twice as bright as a single star. 

\section*{Acknowledgements}
We thank the anonymous referee for a helpful report.
We thank Eliot Quataert, Chao Liu, David Hogg, and Andy Gould for useful conversations and comments. 
K.E. acknowledges support from the SFB 881 program (A3), a Berkeley Fellowship, a Hellman award for graduate study, and an NSF graduate research fellowship. 
H.W.R. received support from the European Research Council under the European Union's Seventh Framework Programme (FP 7) ERC Grant Agreement n. [321035]. 
Y.S.T is supported by the Australian Research Council Discovery Program DP160103747.
D.R.W. is supported by a fellowship from the Alfred P. Sloan Foundation. 
C.C. acknowledges support from NASA grant NNX15AK14G, NSF grant AST-1313280, and the Packard Foundation.

\bibliographystyle{mnras}

\label{lastpage}
\end{document}